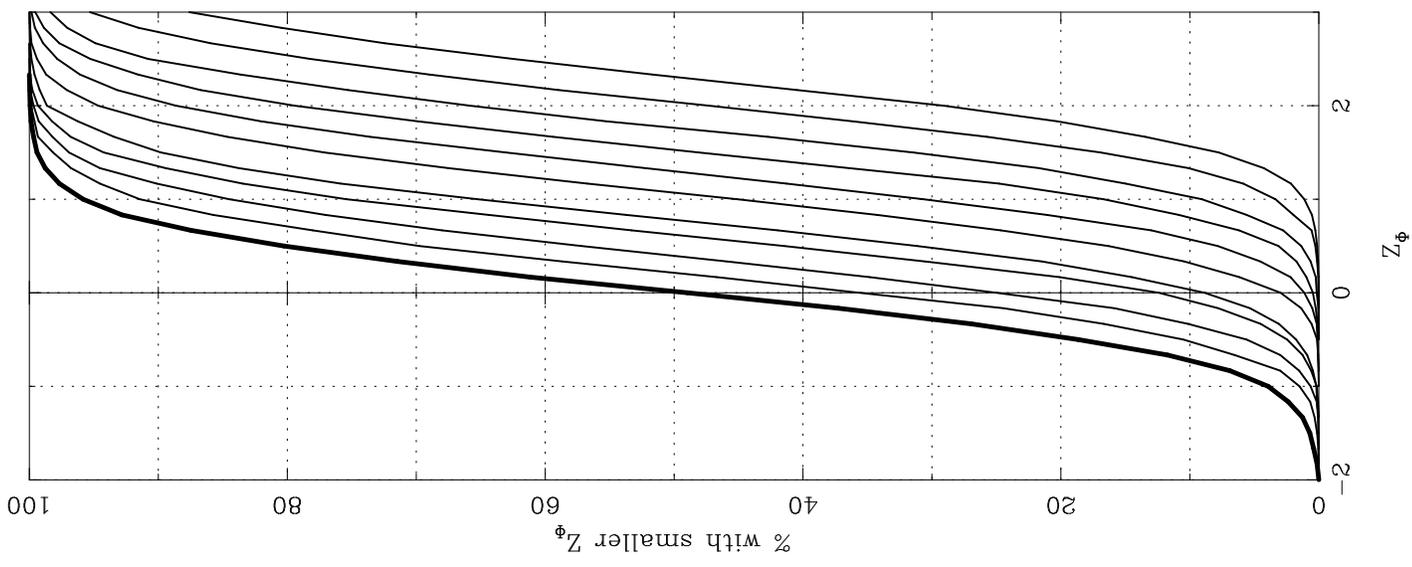

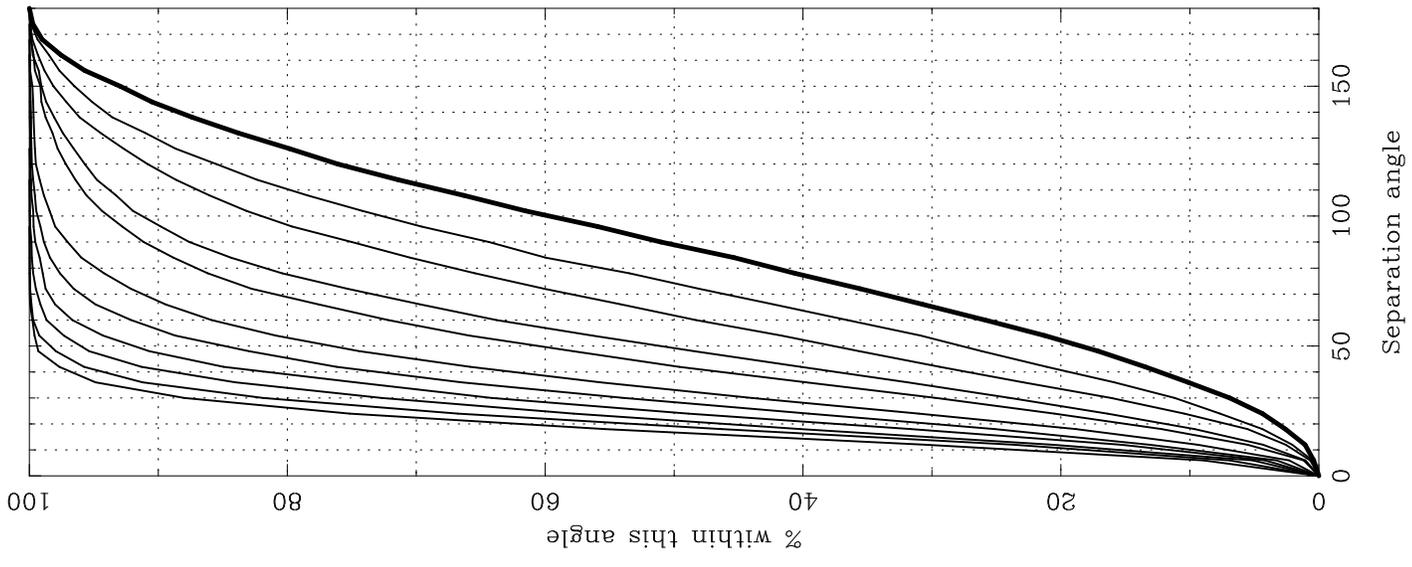

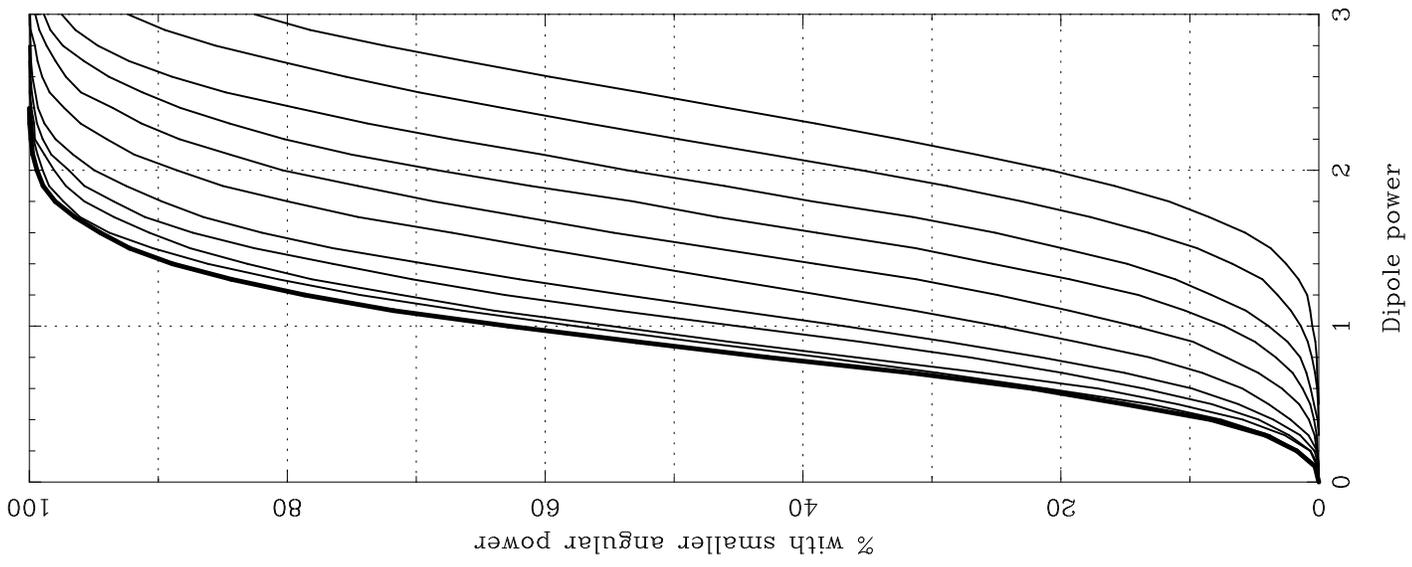

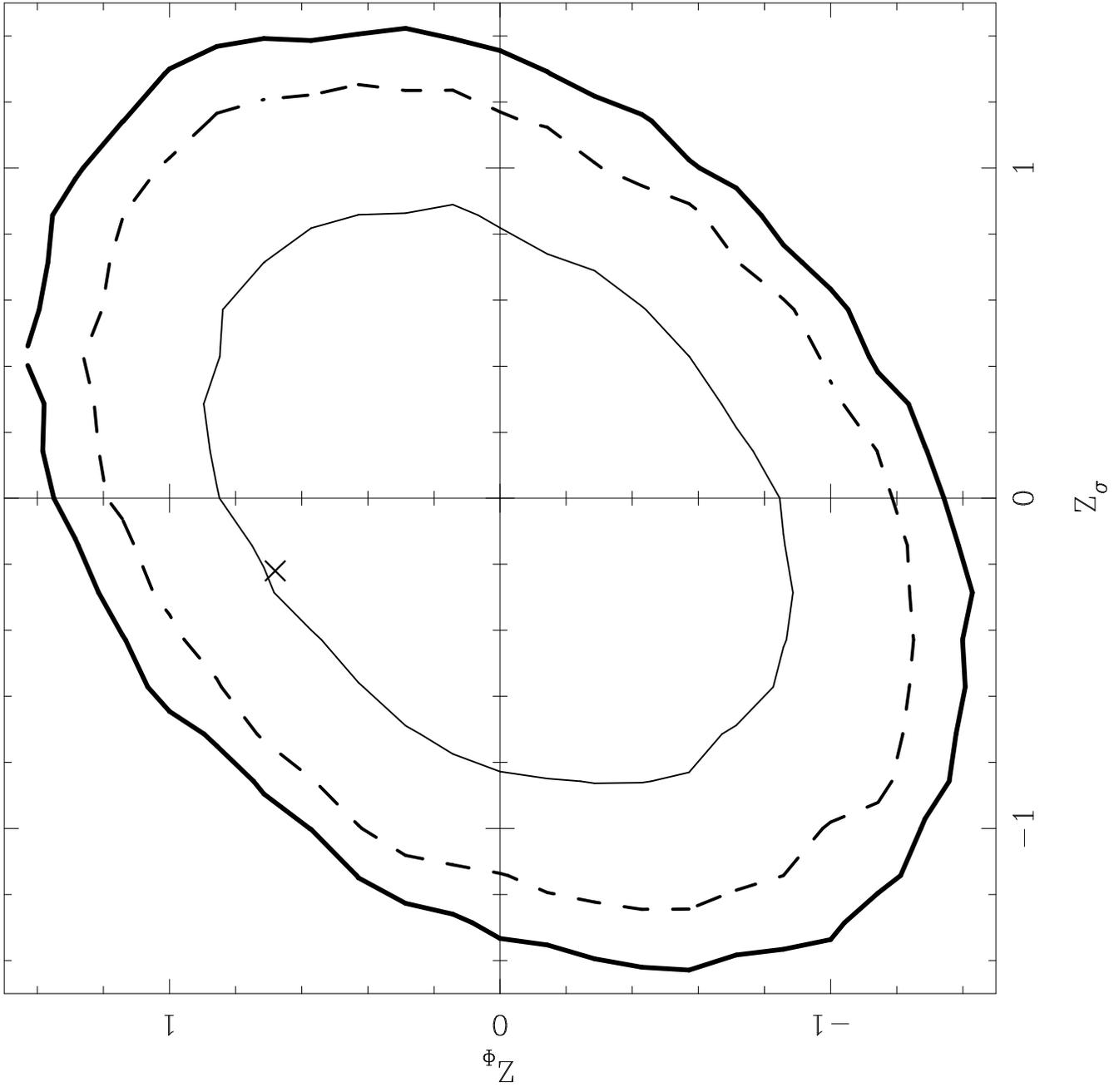

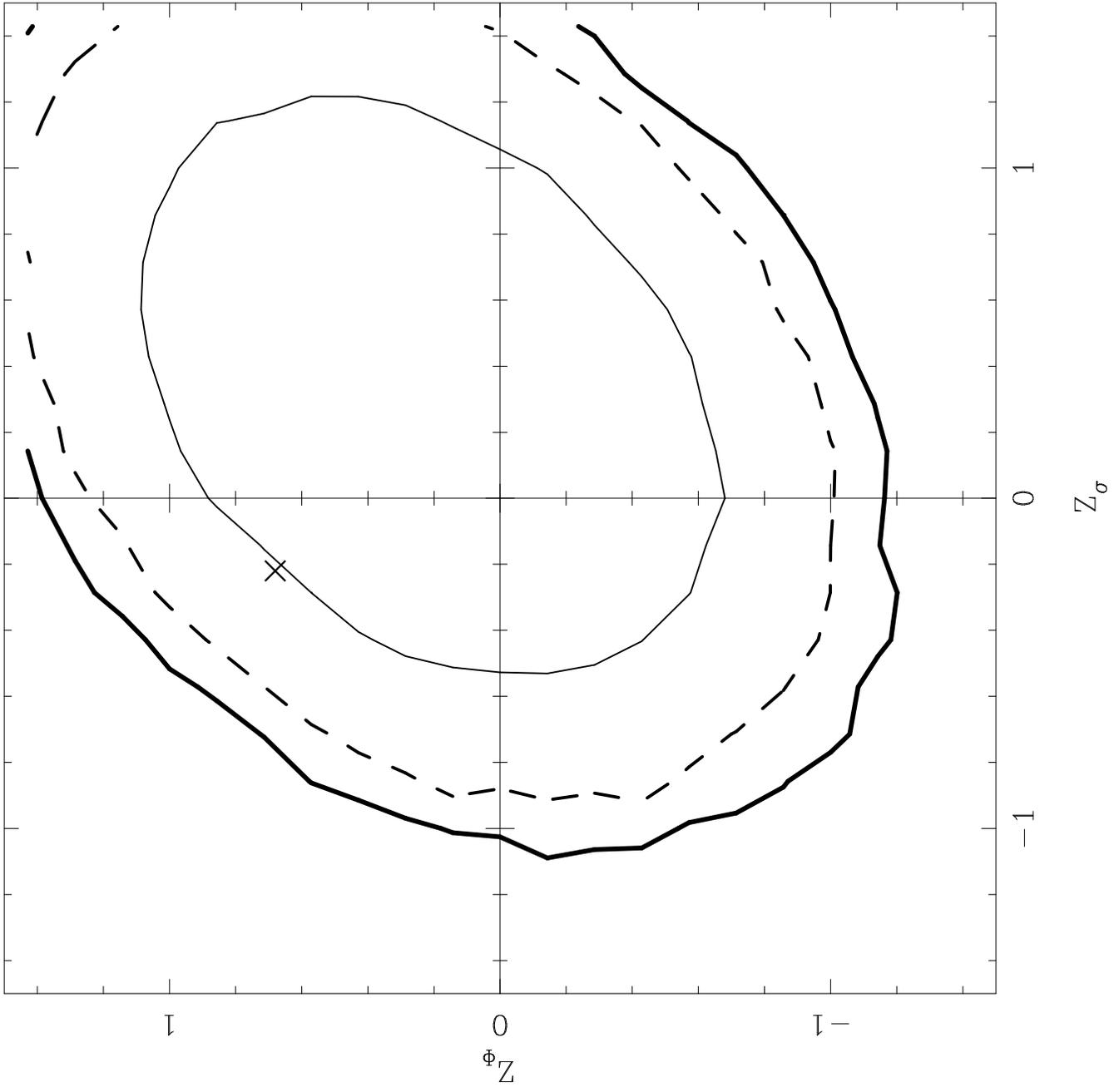

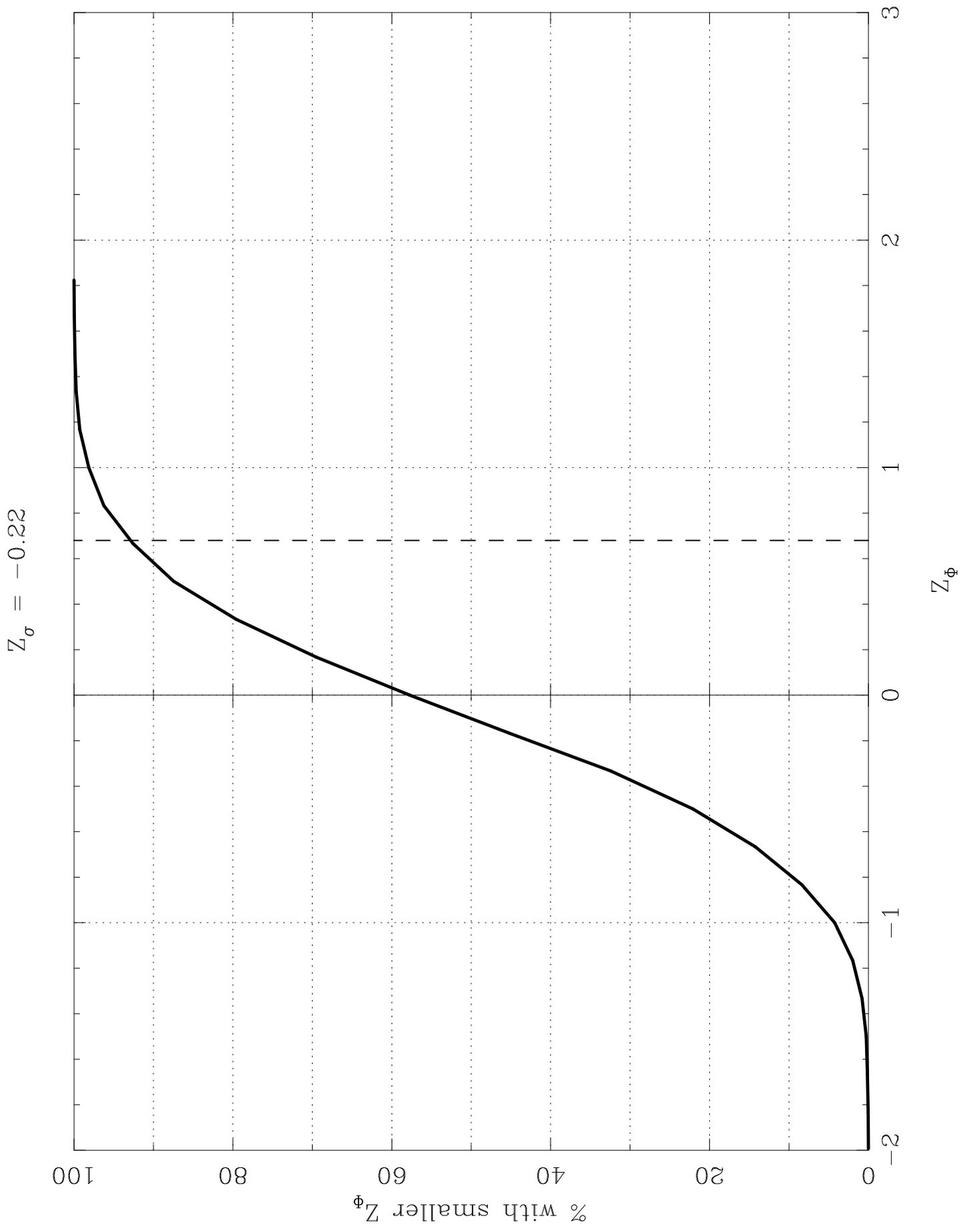

# ON THE MEASUREMENT OF A COSMOLOGICAL DIPOLE IN THE PHOTON NUMBER COUNTS OF GAMMA-RAY BURSTS

C. A. SCHARF [1], K. JAHODA & E. BOLDT

NASA/Goddard Space Flight Center, Code 666, Greenbelt, MD 20771

## ABSTRACT

If $\gamma$-ray bursts are cosmological or in a halo distribution their properties are expected to be isotropic (at least to $1^{st}$ order). However, our motion with respect to the burst parent population (whose proper frame is expected to be that of the Cosmic Microwave Background (CMB), or that of a static halo) will cause a dipole effect in the distribution of bursts *and* in their photon number counts (together termed a Compton-Getting effect).

We argue that the photon number count information is necessary to distinguish a genuine Compton-Getting effect from some other anisotropy and to fully test the proper frame isotropy of the bursts. Using Monte-Carlo simulations we obtain probability distributions for the statistics of dipole alignment, angular power and dipole aligned component. We demonstrate the agreement expected between number distribution and photon count distribution dipoles in the presence of noise. It is estimated that of the order $10^4$ bursts would be necessary to constrain a dipole effect of 1%. However we can test the consistency of number and photon count distributions for a catalogue of any size. Using the 2B catalogue (Meegan et al 1994) of bursts observed with the COMPTON/BATSE instrument (Fishman et al 1989) (in the energy band 20-50$keV$), and the dipole determined from the CMB, we find the surprising result that although the number weighted distribution is consistent with isotropy, the fluence weighted dipole has a correlation with the CMB dipole that has a probability of occuring only 10% of the time for an isotropic photon distribution. Furthermore, the photon and number dipoles are inconsistent under the hypothesis of isotropy, at the $2\sigma$ level. Taken together this could be an indication that a non-negligible fraction of $\gamma$-ray bursts originate in the local, anisotropic universe.

---

[1]NRC Research Associate

These results suggest that future analyses of the angular distribution of $\gamma$-ray bursts should include both photon count and number weighting and that larger catalogues should be used to test the robustness of the apparent inconsistency with isotropy found here.

*Subject headings:* cosmology:observations — gamma ray bursts:bursts — large scale structure:

## 1. Introduction

Measurements of our motion with respect to the proper frame defined by the Cosmic Microwave Background (CMB) are of great importance in the determination of large scale dynamics and measurements of the global density parameter (e.g. Lahav et al 1988, Strauss et al 1992). However, the clear identification of a distant population of objects that can be considered to be at rest in the CMB frame has yet to be made (e.g. Lauer & Postman 1994). The distinct nature of a Compton-Getting (see below) dipole effect due to our motion with respect to the radiation field of such a population offers the possibility of making this identification. An agreement between this dipole anisotropy and the CMB dipole would strongly support the cosmological origin of the source(s) of the radiation field in question[2]. A Compton-Getting dipole effect *not* correlated with the CMB dipole would indicate a different, and possibly local origin for the radiation field. In the case of $\gamma$-ray bursts it is this diagnostic property of the Compton-Getting effect that motivates this present work. We are principally concerned with demonstrating that the photon number count information is necessary in correctly identifying and measuring a Compton-Getting effect in either cosmologically distributed $\gamma$-ray bursts or static halo model bursts. In addition we show that the measurement of the angular distribution of bursts is an insufficient test of isotropy. A truly isotropic population should exhibit isotropy *in all intrinsic properties*, and all such distributions should agree within the given noise.

Our analysis of the photon number count dipole is based on two simple assumptions. Firstly, we assume that if the bursts are indeed cosmological (e.g. arise in a volume of space out to $z \simeq 1$) they arise in a frame at rest with respect to the CMB (this assumption identifies a preferred direction and restricts our search managably in the presence of shot noise). If the bursts arise from a local halo we assume that this population is at rest with respect to the Galactic centre (an assumption which may be inconsistent with some halo models). Secondly, we assume that we can apply the ergodic hypothesis and treat a catalogue of bursts collected over a period of years, and which is instantaneously extremely non-uniform, as a fair and instantaneous sample of the parent population of burst objects. Note that previous tests of isotropy in the burst number distribution make the same assumptions.

Maoz (1994) has calculated the enhancement of burst number counts in the direction of motion with respect to the CMB. This special relativistic effect is a combination

---

[2]Note that the discovery of an intrinsic dipole distribution in a population of high$-z$ objects, while unexpected, would be extremely significant and call into question our interpretation of the CMB dipole and our models of large-scale structure

of aberration, burst event rate and the 'selection function' of the BATSE experiment. We consider here the additional dipole effect expected in the photon number counts of $\gamma$-ray bursts (a spectral effect). The combination of a number count and a photon count anisotropy is hereafter referred to as the Compton-Getting effect[3].

There is an important distinction between a dipole observed to be consistent with the Sun's motion with respect to the CMB ($370 \pm 10 \text{kms}^{-1}$ towards $l = 265^\circ$, $b = 48^\circ$) and one consistent with either the motion of the Sun with respect to the Galaxy ($238 \pm 6 \text{kms}^{-1}$ towards $l = 89^\circ$, b$=2^\circ$) or the Local Group ($622 \pm 20 \text{kms}^{-1}$ towards $l = 277^\circ$, $b = 30^\circ$) in the CMB frame (Smoot et al 1991). The first two directions might be interpreted as evidence for either a cosmological or halo origin of $\gamma$-ray bursts respectively. The third direction might indicate that the observed dipole is due to intrinsic large scale structures and that a significant fraction of $\gamma$-ray bursts originate in the more local universe. Indeed, only in the isotropic cosmological and halo models of $\gamma$-ray bursts would we expect to be able to accurately measure a genuine Compton-Getting effect (in the halo model this requires a static distribution).

In the cosmological model there is considerable uncertainty in the distribution with redshift of a burst parent population (see section 2 below). It might not be unreasonable therefore to expect $\sim 1\%$ of bursts to originate from a more local volume (to $z \sim 0.01 - 0.02$). From optical, infrared and X-ray studies (e.g. Scharf et al 1992, Miyaji & Boldt 1990) we know that this local volume is significantly anisotropic in terms of the distribution of luminous matter. Since the size of the relativistic anisotropy is also expected to be of the order $\sim 1\%$ (see below) it is apparent that the number count dipole could be biased by the local anisotropies. The angular distribution of a property such as fluence might also reflect the local structure (since higher fluence sources are likely to be nearer). However, by choosing subsets of the data according to fluence one could significantly reduce this bias. We do not address this problem directly in the present work, but note that if our motion with respect to the CMB frame is due to a locally anisotropic mass distribution then the Compton-Getting dipole is expected to be positively correlated with the frame-independent local anisotropy (assuming light traces mass).

For the sake of clarity in demonstrating our approach to this problem we consider below the hypothesis of a cosmological origin of $\gamma$-ray bursts, although much of the discussion of the Compton-Getting effect and its detection could be equally well applied to a static halo

---

[3]A note of historical interest: Compton and Getting (1935) advocated measurements of the flux of cosmic rays as a function of direction, as a means to determine their galactic or extra-galactic origin. Their choice of extragalactic rest frame was that of the globular clusters (a halo population) and the Suns motion with respect to this frame was therefore close to that of its rotation about the Galactic centre.

population.

The spectral effect will be important if the bursts have steep photon indices (which are not equal to unity). Although a variety of spectral forms are observed (e.g. Schaefer et al 1992) the spectral (and temporal) variability among the bursts is initially assumed to be uncorrelated with direction (our second assumption, see above). Our motion will shift the BATSE energy band down(up) in the burst rest frame for blue(red) shifted bursts. For a photon index of $\Gamma > 1$ (where we define the photon index as $\Gamma$ and the number distribution with energy $E$ is $n(E) \propto E^{-\Gamma}$) this will cause the observation of systematically larger fluxes in the direction of motion.

Schaefer et al (1992) find that a typical set of bursts are well fitted by power law photon number spectra, with indices between $+1.36$ and $+2.29$ and that several apparently have breaks in the power law to steeper spectra for energies $\gtrsim 400 keV$. Band et al (1993) find a wider distribution of spectral shapes and include an exponential $(\exp(-E/E_0))$ term at lower energies, as well as breaks at higher energies. We predict a larger spectral effect in the 3rd and 4th channels ($100$-$300 keV$ and $> 300 keV$) due to the steeper photon index. However, as discussed in section 4 below, the broader distribution of higher energy fluences adds shot noise faster than the steeper spectra increase the signal.

The Compton-Getting dipole observed in the CMB corresponds to a Solar system velocity of $370 \pm 10 \mathrm{kms}^{-1}$ (Smoot et al 1992); for energies $\lesssim 400 keV$ this translates to an expected total Compton-Getting effect of the order $1\%$ in burst photon number counts (see section 2 below) under the assumption of $\Gamma \simeq 2$. By weighting the observed burst distribution with an estimate of the photon number count we implicitly fold in the luminosity and duration distributions and the evolution and spatial distribution with $z$. These amount to extra scatter in the resulting distributions of the expected observations (see section 4 below) but offer the possibility of correctly identifying a Compton-Getting effect and testing the proper frame isotropy of the bursts.

In section 2 below we briefly reconsider definitions of the Compton-Getting effect. In section 3 we describe the statistics used here; in section 4 we we describe a set of Monte-Carlo simulations used to determine the significance of such measurements. In section 5 we present the results of applying these tests to a catalogue of 410 BATSE detected bursts with measured fluences and durations. In section 6 the results presented in this work and the future prospects of measuring the Compton-Getting effect are discussed.

## 2. The Compton-Getting effect

In the case of $\gamma$-ray bursts the Compton-Getting effect can be viewed as two multiplicative effects. The first, as derived by Maoz (1994), is the anisotropy in burst *locations*, due to aberration, event rate and the instrumental selection function. The second is manifest if the actual photon number count for each burst is used as a weight. This additional effect is simply due to the red/blue shifting of the burst spectra and the subsequent (for a non-unity photon index) change in detected photon numbers in a given instrumental energy bandpass.

Although not always directly referred to as such, the Compton-Getting effect is responsible for the temperature dipole in the CMB (e.g. Peebles & Wilkinson 1968, Forman 1970) and should be responsible for dipole anisotropies in cosmic ray flux, the diffuse X-ray background and any other isotropic source population or radiation field with respect to which we have a non-zero relative velocity.

The number of bursts per steradian $\sigma$ in a direction $\hat{\mathbf{r}}$ given our direction of motion $\hat{\mathbf{r}}_V$ is therefore (in a given time interval):

$$\sigma(\hat{\mathbf{r}}) \quad \propto \quad (1 + 2\beta \ \hat{\mathbf{r}} \cdot \hat{\mathbf{r}}_V)(1 + \beta \ \hat{\mathbf{r}} \cdot \hat{\mathbf{r}}_V)(1 + K\beta \ \hat{\mathbf{r}} \cdot \hat{\mathbf{r}}_V) \ , \tag{1}$$

where $\beta = v/c$ ($\beta << 1$) and a factor $(1 + 2\beta \ \hat{\mathbf{r}} \cdot \hat{\mathbf{r}}_V)$ comes from aberration, a factor $(1 + \beta \ \hat{\mathbf{r}} \cdot \hat{\mathbf{r}}_V)$ from event rate, and a factor $(1 + K\beta \ \hat{\mathbf{r}} \cdot \hat{\mathbf{r}}_V)$ from the instrumental selection function and assumptions about the population evolution/distribution with $z$ (Maoz 1994).

For an average burst photon number spectrum $n(E) = A \ E^{-\Gamma} \ (photons/keV/cm^2)$ the observed number of photons per $cm^2$ per steradian is then:

$$\Phi(\hat{\mathbf{r}}) \quad \propto \quad (1 + (\Gamma - 1)\beta \ \hat{\mathbf{r}} \cdot \hat{\mathbf{r}}_V) \ \sigma(\hat{\mathbf{r}}) \quad . \tag{2}$$

The instrumental efficiency (effective area) is in reality, energy dependent, however, to first order we can ignore this additional effect. Note that this spectral effect vanishes for $\Gamma = 1$, where the change in bandwidth is exactly compensated for by the spectral shift.

Since $1.0 \leq \Gamma \leq 2.5$ (Schaefer et al 1992) and $K \simeq 1$ for most reasonable models (Maoz 1994) with a Solar system velocity of $\sim 300 - 400 km s^{-1}$ the fore-aft asymmetry should have a magnitude of $\simeq 1\%$. We henceforth denote the overall amplitude of the Compton-Getting effect as $D$, where (for $\beta << 1$):

$$\Phi(\hat{\mathbf{r}}) \quad \propto \quad (1 + D \ \hat{\mathbf{r}} \cdot \hat{\mathbf{r}}_V) \quad , \tag{3}$$

i.e. $D \approx (\Gamma + 2 + K)\beta$ .

The additional photon number count information in $\Phi$ is necessary to distinguish a genuine, motion induced, Compton-Getting effect from some other anisotropy due to intrinsic structure in the burst distribution. Firstly, an important test can be made by comparing the dipole axis seen in $\Phi$ with that of $\sigma$. In the absence of noise they should show exact agreement; in the presence of noise they should be distributed about each other with a scatter determined by the signal/noise ratio (see section 4 below). Secondly, by taking the ratio of the functions $\Phi/\sigma$ a direct measurement of our peculiar motion (given an estimate of $\Gamma$) can, in principle, be made without any knowledge of the relatively uncertain parameter $K$. In reality, with discrete data, this ratio is ill defined unless we smooth the data in some way. By writing $\sigma$ and $\Phi$ as finite (i.e. to some angular resolution) sum over spherical harmonics it can, in principle, be determined. Here however, we are primarily concerned with simply evaluating our ability to detect/rule out a Compton-Getting effect and test the proper frame isotropy of the bursts. The following section discusses the measurements used.

## 3. Observable quantities

The BATSE catalogue provides estimated burst fluences. Here we treat the fluences directly as photon number counts in a given band (equivalent to assuming that a constant spectrum has been used to convert all burst number counts to $ergs/cm^2$). The dipole is then simply determined as a sum over bursts:

$$a_{1,m} = \sum_i w_i Y^*_{1,m}(\theta_i, \phi_i) \,, \qquad (4)$$

where the $Y$ are the orthonormal set of spherical harmonics and $a_{1,m}$ are complex coefficients. The $w_i$ are the weights assigned to each burst, consisting of an exposure correction (due to occultation by the Earth leading to non-uniform burst detection) and a fluence: $w_i = (1/t^{ex}_i)(f_i)$, where $t^{ex}$ is the published fraction of total flight time in a given equatorial band in the 1st+2nd year (2B) BATSE data and $f_i$ is the fluence. As demonstrated below the exposure correction introduces only a small additional scatter in the results. In the case of number weighting we set $w_i = 1/t^{ex}_i$.

The fluence $f$ (units of $ergs/cm^{-2}$) is taken from one of four bands (channel 1: 20-50$keV$, channel 2: 50-100$keV$, channel 3: 100-300$keV$ and channel 4: $> 300keV$). The errors on these measured fluences are typically of the order $< 10\%$. The positional errors of the observed bursts are typically $\sim 4^o$ and $\sim 13^o$ for bright and dim bursts respectively. As discussed below these sources of uncertainty are small compared to the shot noise uncertainties.

An estimator of the dipole angular power (normalized by the mean shot noise $(\frac{1}{4\pi}\sum_i w_i^2)$) is:

$$C = \left[ \frac{4\pi}{3\sum_i w_i^2} \sum_{m=-1,0,1} |a_{1,m}|^2 \right]^{1/2} . \qquad (5)$$

The mean of this distribution is unity (and the median somewhat lower) when there is no signal, since the observed $a_{1,m}$ will just be noise. The dipole direction is simply the direction of the 3-vector formed from the complex coefficients $a_{1,m}$:

$$\mathbf{r} = (a_{1,-1} - a_{1,+1}, \quad \frac{1}{\mathbf{i}}[a_{1,+1} + a_{1,-1}], \quad \sqrt{2}a_{1,0}) \qquad (6)$$

where, in Galactic coordinates the polar angle $\theta = b$ and the azimuthal $\phi = l$.

The question of what constitutes a significant dipole is therefore two-fold. It is a combination of the angular power ($C$) and the degree of alignment with the expected dipole direction. In the following sections the component of the observed dipole power that lies along the axis of the expected dipole is used as the preferred statistical measure since it combines both quantities. This is simply

$$Z_{comp} = C_{observed} \; (\hat{\mathbf{r}}_{observed} \; \cdot \; \hat{\mathbf{r}}_{expected}) \quad , \qquad (7)$$

which has the advantage over a simple angular power measurement of including the additional prior information of the expected dipole direction. We hereafter refer to the $Z_{comp}$ of the number and fluence weighted distributions as $Z_\sigma$ and $Z_\Phi$ respectively.

We note that the statistic presented here is essentially that used by Briggs (1993), albeit cast in a different form more commonly seen in analyses of extra-galactic catalogues (e.g. Scharf et al 1992).

## 4. Monte Carlo simulations

Given our lack of detailed knowledge of the parent distributions of luminosities, spectra and distances of gamma-ray bursts we to turn to Monte Carlo simulations in order to determine the probability distributions of dipole angular power, alignment and aligned component ($Z_{comp}$). Catalogues of random positions are generated and fluences drawn from the BATSE catalogue are assigned to these random positions.

First we address the question of what properties of a catalogue (such as size) are desirable in order to make a robust estimate of the Compton-Getting effect. For each Monte-Carlo realization a $4\pi$ steradian catalogue of angular positions is generated with the burst location probability varying on the sphere according to the chosen observers velocity. Each position is then assigned a set of burst properties drawn at random from the 410 sets of catalogued burst parameters (see section 5). The 'observed' fluences/fluxes are then perturbed by the spectral effect. This anisotropy and the positional anisotropy are constructed to have a total Compton-Getting amplitude of $D$. (For the results presented here we fix the ratio of spectral to positional anisotropy amplitudes to be $\sim 1 : 2$. Varying this ratio does not alter the resulting probability distributions at any significant level, since we are dominated by shot noise and the breadth of the fluence distribution).

For each value of $D$ the probability distribution of the observed angular power $C$, observed alignment with the Compton-Getting dipole and the observed $Z_\sigma$ or $Z_\Phi$ is built up from the Monte-Carlo simulations. Figure 1 illustrates the nature of the probability distributions expected for a catalogue of 10,000 fluence weighted bursts (using channel 1 fluences). The cumulative probability curves are presented for the fluence dipole, i.e. for a given observation the fraction of observers who would make a measurement of this magnitude or less is plotted against (a) the angular power, (b) alignment and (c) $Z_\Phi$. It should be noted that although we present results for Compton-Getting effects as large as 20% we have *not* included second order relativistic corrections. These results should therefore be considered only as a general guide, since realistically we are only concerned with an effect of the order 1%.

The instrumental exposure effect has only a small influence on the dipole measurements (see section 5) and is not included in the simulations. In reality it will add a small component to the noise in the measurements. Similarly, positional errors will introduce uncertainties of at least an order of magnitude less than the uncertainties due to a discrete (Poisson) distribution, irrespective of the number of objects considered here.

As might be expected, the folding in of a spread of fluences and spectra considerably decreases the sensitivity of the absolute dipole measurement in $\Phi$. Plots similar to Figure 1, but created with a $\delta$-function fluence distribution (where $Z_\Phi = Z_\sigma$) indicate that the Compton-Getting effect can be constrained to be $1\% \lesssim D \lesssim 5\%$ (at 90% limits). An observer measuring a $Z_\Phi$ of 1 in our more realistic $20 - 50 keV$ channel could only constrain the effect to $1\% \lesssim D \lesssim 10\%$ (90% limits).

The null hypothesis ($D = 0$) distribution (heaviest curve in all plots) is identical for any size burst catalogue (since we normalize by the mean shot noise), though for a smaller catalogue (such as 410 bursts) the effect of increasing $D$ is far less significant. From these plots we can see that both the observed dipole angular power and direction are useful indicators, and that the aligned component $Z_{comp}$ effectively combines these quantities. The scatter in observed $Z_\Phi$ is constant for a given catalogue, but the mean value increases with increased $D$.

A comparison of the results obtained using the fluences in the different channels demonstrates that similar results are obtained with the higher energy bands. However, the spread in observed fluences increases with the energy band. As a simple measure of this scatter we evaluate the variance of the distribution of fluence divided by mean fluence for each channel (this is just a convenient measure of the width of the decidedly non-Gaussian distributions). In Table 1 the normalized variances are presented. There is a systematic increase with energy, with channel 4 fluences having a relative variance approximately 5 times that of channel 1 fluences. Clearly the ability to constrain the magnitude of a Compton-Getting effect will be impaired in the higher energy bands, especially the highest, due to this extra scatter.

An additional test was performed in which the burst duration data was incorporated. Using the period during which 90% of the burst counts were obtained (the so-called 't90' time) we can determine a crude measure of flux and weight the bursts by this. There was no significant gain in doing this, i.e. to the sensitivity of this statistic there is no correlation between duration and fluence which conspires to narrow the spread of fluxes compared to that of fluences.

From these results it is apparent that *at least* 10,000 bursts would be necessary to quantify a $\sim 1\%$ Compton-Getting dipole with some confidence. However, we can still address the question of whether the number and fluence weighted distributions are consistent under a given hypothesis.

A fully rigorous test for the Compton-Getting dipole would be to compare the observations with the joint probability distribution of the expected $\sigma$ and $\Phi$ dipoles (determined by a total of 6 numbers). To simplify the problem we choose a preferred direction (in the case of the real data we choose the CMB dipole direction) and consider the joint distribution of $Z_\sigma$ and $Z_\Phi$ (2 numbers). To illustrate this the probability distribution of $Z_\sigma$ and $Z_\Phi$ for 410 bursts with zero intrinsic dipole ($D = 0$, the null hypothesis) from 200,000 Monte-Carlo realizations is plotted in Figure 2a. Defining areas which contain a certain fraction of the population is somewhat arbitrary. The contouring chosen here is along lines of equi-density and the contours enclose 68%,90% and 95% of the total population (see Figure 2a). The contours were constructed by ranking the counts in bins and then adding them in descending size until a particular population fraction was reached. The implicit correlation between $\sigma$ and $\Phi$ due to common burst positions is seen in the orientation of the major axis of the distribution. In Figure 2b the results for a catalogue of 410 bursts but with $D = 10\%$ are shown to demonstrate the effect of an intrinsic dipole, the major effect of which is to shift the centroid of the distribution.

In the following section we compare real burst data with these expectations.

## 5. Results from 410 BATSE bursts

Although a small dataset and undoubtedly prone to instrumental biases not considered here we apply the dipole tests to the existing BATSE public burst catalogue obtained from the *Compton GRO* public domain, available through the Compton Science Support Center at NASA Goddard Space Flight Center.

The 1st + 2nd year publicly available (2B) BATSE catalogue contains 585 burst positions; of these only 410 have both fluence and duration information. Since we compare our fluence weighted results to a flux weighted dipole, we limit ourselves to this subset. Our investigation of possible biases introduced by any incomplete burst data does not reveal any significant trend in the angular distribution of these bursts. The majority of the data gaps occur in the 2nd year catalogue; below we perform a simple comparison of 1st and 2nd years and do not find any strong anti-correlation. The effect of the exposure correction ($1/t^{ex}$) is small and does not alter the results significantly (see Table 2). An additional check was made of the higher order harmonics (e.g. quadrupole and higher) for any possible instrumental effects. To the sensitivity obtainable from this dataset there is no significant higher order structure.

In Table 2 the results for different channels are shown with respect to the direction of the Sun's motion in the CMB frame; ($265°$, $48°$) in Galactic coordinates (Smoot et al 1992). For the 20-50$keV$ channel the $\Phi$ dipoles from 1st and 2nd year data separately would be this aligned (or more) $\sim 43\%$ of the time if they were two random vectors.

Constructing the appropriate probability distributions as described above we find that the observed $Z_\Phi$ (with respect to the CMB dipole) or *greater* would only occur with a 10% probability for the null hypothesis of isotropy. For a 40% Compton-Getting effect ($D = 0.4$) the observed $Z_\Phi$ or *less* would occur with a 10% probability. Therefore we can state that the observed bounds are $0\% \lesssim D \lesssim 40\%$ (90% limits).

As a confirmation of the previously seen isotropy in the number weighted distribution of bursts (Briggs et al 1993) the bottom row of Table 2 shows the results for a pure number weighted (but exposure corrected) dipole estimation for the 410 bursts used here. The resulting $Z_\sigma$ (with respect to to CMB dipole) or greater would occur $\sim 65\%$ of the time in the null hypothesis case of no number density dipole in the CMB dipole direction. The joint probability of $Z_\sigma$ and $Z_\Phi$ lying outside the population defined by the equi-density at $Z_\sigma = -0.22$ and $Z_\Phi = 0.68$ in the null hypothesis (Figure 2a) is $\sim 32\%$. In Figure 3 the cumulative probability of $Z_\Phi$ given the observed $Z_\sigma$ is plotted for the null hypothesis. The observed $Z_\Phi$ or greater would be expected only $\sim 6\%$ of the time (assuming the probability distribution is symmetric about the mean this corresponds to a $\sim 2\sigma$ deviation from the mean).

The possibility of a few high fluence bursts being responsible for this result was checked by excluding those bursts above half the maximum observed fluence (a total of 20 bursts). The resulting $Z_\Phi$ was 0.70, entirely consistent with the result for the complete dataset, given the level of shot noise. The effect of the exposure correction is small; leaving it out we see (Table 2) that the null hypothesis is still rejected at the 85% level for the channel 1 $Z_\Phi$. Similar deviations are obtained for the other channels.

For the sake of completeness we also compare the observed fluence dipole with the direction of motion of the Sun with respect to the Galaxy (halo) and the Local Group with respect to the CMB. The $Z_\Phi$ with respect to the Suns motion around the Galaxy is -0.6; this value or less would occur $\sim 20\%$ of the time in the null hypothesis. For channels 1 and 2 there is however a small increase (to $\sim 5\%$) in the significance of $Z_\Phi$ using the direction of the Local Group motion. In the null hypothesis the observed channel 1 $Z_\Phi$ or greater would occur 5% of the time, for channel 2 $Z_\Phi$ or greater would occur 10% of the time.

The results for the 410 observed bursts can be summarised as follows (with respect to the CMB dipole). The observed $Z_\sigma$ is consistent with a random (isotropic) distribution, the observed $Z_\Phi$ or greater would only be seen $\sim 10\%$ of the time in the null hypothesis case. However, for the observed number distribution (parameterised by $Z_\sigma$) then the observed $Z_\Phi$ or greater would only be seen $\sim 6\%$ of the time (a $\sim 2\sigma$ result assuming a symmetric probability distribution), suggesting an inconsistency between the number and fluence weighted distributions under the null hypothesis of an isotropic distribution. The use of future burst catalogues will determine the robustness of this result.

While these results are surprising it is clear from Table 2 that the results from other channels are less dramatic. The observed trend of decreasing $Z_\Phi$ with increasing energy is consistent with the greatly increased scatter of burst fluences with increasing energy band (see section 4, Table 1).

## 6. Discussion

We have presented here an evaluation of the Compton-Getting effect that might be expected in the photon number counts from $\gamma$-ray bursts. The additional information in photon number counts allows both a self-consistency check of the effect and, in principle, a direct measurement of our motion with respect to the bursts with less bias than is present in the number distribution.

A positive measurement correlated with the CMB motion vector would provide very strong evidence for the extra-galactic and high-$z$ origin of $\gamma$-ray bursts. Similarly a positive measurement correlated with the Suns motion relative to the Galaxy would support a static halo origin. However, a catalogue of at least 10,000 bursts does seem necessary in order to quantify a $\sim 1\%$ Compton-Getting dipole. The agreement between number weighted and fluence weighted dipoles can be determined by considering the joint probability of $Z_\sigma$ and $Z_\Phi$ and the conditional probability of $Z_\Phi$ given $Z_\sigma$.

A measurement from 410 BATSE bursts is presented; given only 410 bursts, on the basis of our Monte-Carlo simulations a 1% Compton-Getting effect could not provide a significant signal to noise. However, we find the surprising result that the fluence and number weighted dipoles are inconsistent for the null hypothesis at roughly the $2\sigma$ level. Furthermore the measured $Z_\Phi$ or greater (with respect to the CMB dipole direction) would only occur 10% of the time for an isotropic fluence distribution. It is particularly interesting that the correlation with the direction of motion of the Local Group is more significant (for the null hypothesis it should occur only $\sim 5\%$ of the time). If we are measuring a real correlation with the Local Group motion vector then this is evidence for a more local origin of the $\gamma$-ray bursts. Together these results could indicate that a non-negligible fraction of $\gamma$-ray bursts originate within the local (anisotropic) universe. There is no evidence of correlation with the direction expected in a halo model.

In conclusion, the use of a burst property such as fluence is necessary in the proper evaluation of deviations from isotropy and in correctly identifying a Compton-Getting effect. We anticipate that future burst catalogues will be sufficiently large that statistically useful subsets could be constructed using a variety of astrophysically interesting selection criteria (e.g. spectral form, temporal width). These more extensive data will allow a detailed investigation of the nature of anisotropies such as that suggested by this preliminary work.

We wish to thank E. Gotthelf for useful discussions. C.S. acknowledges the National Research Council for their support through a Research Associateship.

| | $20-50 keV$ | 50-100$keV$ | 100-300$keV$ | $> 300 keV$ |
|---|---|---|---|---|
| Variance | 2.9 | 5.4 | 6.0 | 14.1 |

Table 1: The variances of the (fluence)/(mean fluence) distributions for the four energy channels used here.

| Fluences | C | Direction $(l, b)$ | Alignment | $Z_{comp}$ |
|---|---|---|---|---|
| 20-50$keV$ (1st+2nd yr) | 1.04 | $(320^o, 23^o)$ | $49^o$ | 0.68 |
| 20-50$keV$ (1st yr) | 0.84 | $(270^o, 29^o)$ | $19^o$ | 0.79 |
| 50-100$keV$ (1st+2nd yr) | 0.85 | $(324^o, 24^o)$ | $52^o$ | 0.52 |
| 100-300$keV$ (1st+2nd yr) | 0.54 | $(7^o, 19^o)$ | $84^o$ | 0.06 |
| $> 300 keV$ (1st+2nd yr) | 0.70 | $(128^o, 19^o)$ | $103^o$ | -0.16 |
| 20-50$keV$ No exposure corrn. | 0.90 | $(328^o, 28^o)$ | $52^o$ | 0.55 |
| | | | | |
| Number weighted (1st+2nd yr) | 0.81 | $(8^o, -10^o)$ | $106^o$ | -0.22 |

Table 2: Fluence weighted results from 410 BATSE bursts with measured positions, fluences and durations. Second row from bottom is fluence weighted, but with no exposure correction. Bottom row is the number weighted result for the same 410 bursts. Alignment is measured with respect to the CMB dipole due to the Sun's motion.

---

**Figure Captions**

**Figure 1:** Cumulative probability distributions over 2000 Monte Carlo simulations of 10,000 bursts with fluences drawn at random from the 410 observed fluences. All results are for the fluence weighted distribution. The heaviest curve in each plot corresponds to the zero dipole case $D = 0\%$, subsequent curves are for $D = 2\%$, 4%, 6% ... Compton-Getting effects. a) fraction with smaller angular power, b) fraction within a given alignment angle and c) the fraction with smaller dipole component along the direction of the true dipole.

**Figure 2:** (a) The joint distribution of $Z_\Phi$ and $Z_\sigma$ for a catalogue of 410 bursts with $D = 0$. Contours levels are along lines of equi-density, the innermost (thin solid line) contour contains 68% of observations, the dashed contour contains 90% and the heavy solid contour contains 95%. The cross symbol at $Z_\sigma = -0.22$, $Z_\Phi = 0.68$ is the result from the real 410 burst catalogue. (b) as for (a) but with $D = 10\%$ to illustrate the results of a Compton-Getting like effect.

**Figure 3:** The cumulative probability distribution of $Z_\Phi$ expected given the observed number weighted distribution in the 2B burst catalogue (with $Z_\sigma = -0.22$). The vertical line (heavy dashed) at $Z_\Phi = 0.68$ corresponds to the observed channel 1 fluence weighted dipole.